\documentclass[reprint,twocolumn,superscriptaddress,prl,showpacs,amsmath,amssymb,aps]{revtex4-1}

\usepackage{natbib}	
\usepackage{graphicx,color,rotating}
\usepackage[latin1]{inputenc}
\usepackage{textcomp}
\usepackage{dcolumn}
\usepackage{bm}     
\usepackage{upgreek}
\usepackage{subdepth}  			
\usepackage[latin1]{inputenc}

\usepackage{array}
\newcolumntype{L}[1]{>{\raggedright\let\newline\\\arraybackslash\hspace{0pt}}m{#1}}
\newcolumntype{C}[1]{>{\centering\let\newline\\\arraybackslash\hspace{0pt}}m{#1}}
\newcolumntype{R}[1]{>{\raggedleft\let\newline\\\arraybackslash\hspace{0pt}}m{#1}}

\usepackage{hyperref}						
\usepackage{xcolor}						
\hypersetup{							
    colorlinks,
    linkcolor={blue!50!black},
    citecolor={blue!30!black},
    urlcolor={blue!80!black}
}
\renewcommand{\vec}[1]{\bm{#1}}
\newcommand{\ee}{\mathrm{e}}
\newcommand{\ii}{\mathrm{i}}

\newcommand{\avr}[1]{\big\langle #1 \big\rangle}

\newcommand{\iot}{{\ii\omega t}}

\newcommand{\pp}{\partial}

\newcommand{\nablabf}{\boldsymbol{\nabla}}

\newcommand{\Lapl}{\nabla^2}

\newcommand{\divop}{\nablabf\cdot}

\newcommand{\fffac}{\vec{f}_\mathrm{ac}}

\newcommand{\gvec}{\vec{g}}

\newcommand{\rrr}{\vec{r}}

\newcommand{\vvv}{\vec{v}}

\newcommand{\zerovec}{\boldsymbol{0}}

\newcommand{\cO}{c_0}

\newcommand{\Eac}{E_\mathrm{ac}}

\newcommand{\pii}{p_{11}}                  

\newcommand{\etaO}{\eta_0}

\newcommand{\nuO}{\nu_0}

\newcommand{\pti}{{\tilde{p}{}}}

\newcommand{\kapO}{\kappa_0}

\newcommand{\pI}{p_1}

\newcommand{\vvvI}{\vvv_1}

\newcommand{\rhoO}{\rho_0}
\newcommand{\rhoOO}{\rho_0^{(0)}}

\newcommand{\rhoI}{\rho_1}

\newcommand{\rhoHat}{\hat{\rho}}

\newcommand{\SIum}{\upmu\textrm{m}}

\newcommand{\SImum}{\textrm{\textmu{}m}}

\newcommand{\SIPa}{\textrm{Pa}}

\newcommand{\SIs}{\textrm{s}}
\newcommand{\SIms}{\textrm{ms}}

\newcommand{\beq}[1]{\begin{equation} \eqlab{#1}}
\newcommand{\eeq}{\end{equation}}
\newcommand{\bsub}{\begin{subequations}}
\newcommand{\esub}{\end{subequations}}
\def\bal#1\eal{\begin{align}#1\end{align}}
\def\bsubal#1\esubal{\bsub \begin{align}#1\end{align} \esub}

\newcommand{\eqlab}[1]{\label{eq:#1}}
\renewcommand{\eqref}[1]{Eq.~(\ref{eq:#1})}

\newcommand{\eqrefnoEq}[1]{(\ref{eq:#1})}
\newcommand{\eqsref}[2]{Eqs.~(\ref{eq:#1}) and~(\ref{eq:#2})}

\newcommand{\figref}[1]{Fig.~\ref{fig:#1}}

\newcommand{\figlab}[1]{\label{fig:#1}}

\newcommand{\Piac}{\bm{\Pi}_\mathrm{ac}}
\newcommand{\sigmabf}{\bm{\sigma}}

\newcommand{\sigmabfI}{\bm{\sigma}^{{}}_1}

\begin{document}

\title{Acoustic Streaming and its Suppression in Inhomogeneous Fluids}

\author{Jonas T. Karlsen}
\email{jonkar@fysik.dtu.dk}
\affiliation{Department of Physics, Technical University of Denmark, DTU Physics Building 309, DK-2800 Kongens Lyngby, Denmark}

\author{Wei Qiu}
\affiliation{Department of Physics, Technical University of Denmark, DTU Physics Building 309, DK-2800 Kongens Lyngby, Denmark}

\author{Per Augustsson}
\affiliation{Department of Biomedical Engineering, Lund University, Ole R\"{o}mers v\"{a}g 3, 22363, Lund, Sweden}

\author{Henrik Bruus}
\email{bruus@fysik.dtu.dk}
\affiliation{Department of Physics, Technical University of Denmark, DTU Physics Building 309, DK-2800 Kongens Lyngby, Denmark}

\date{23 July 2017}

\begin{abstract}
We present a theoretical and experimental study of boundary-driven acoustic streaming in an inhomogeneous fluid with variations in density and compressibility. In a homogeneous fluid this streaming results from dissipation in the boundary layers (Rayleigh streaming). We show that in an inhomogeneous fluid, an additional non-dissipative force density acts on the fluid to stabilize particular inhomogeneity configurations, which markedly alters and even suppresses the streaming flows. Our theoretical and numerical analysis of the phenomenon is supported by ultrasound experiments performed with inhomogeneous aqueous iodixanol solutions in a glass-silicon microchip.
\end{abstract}




\maketitle


Acoustic streaming is the steady vortical flow that accompanies the propagation of acoustic waves in viscous fluids. This ubiquitous phenomenon~\cite{Squires2005, Wiklund2012}, studied as early as 1831 by Faraday observing the motion of fine-grained powder in the air above a vibrating Chladni plate~\cite{Faraday1831}, is driven by a non-zero divergence in the nonlinear momentum-flux-density tensor. In a homogeneous fluid, such a divergence is caused by dissipation of acoustic energy by one of two mechanisms. One mechanism is dissipation in the thin boundary layers that emerge near walls in order to match the acoustic fluid velocity with the velocity of the boundary. The resulting streaming, called boundary-driven Rayleigh streaming~\cite{LordRayleigh1884, Schlichting1932}, is typically observed in standing wave fields near walls~\cite{Muller2013} or suspended objects~\cite{Tho2007}. The other mechanism is the attenuation of acoustic waves in the bulk of the fluid, which produces streaming known as bulk-driven Eckart streaming~\cite{Eckart1948} (or the "quartz wind"), typically observed in systems much larger than the wavelength~\cite{Riaud2017a}. Both cases have been extensively studied theoretically~\cite{Nyborg1953a, Nyborg1958, Lighthill1978, Riley2001}, and the phenomenon has continued to attract attention due its importance in processes related to thermoacoustic engines~\cite{Swift1988, Hamilton2003, Hamilton2003a}, ultrasound contrast agents, sonoporation, and drug delivery~\cite{Doinikov2010, Wu2008, Marmottant2003}, and the manipulation of particles and cells in microscale acoustofluidics~\cite{Bruus2011c, Friend2011, Barnkob2012a, Hammarstrom2012, Collins2015, Marin2015, Hahn2015a, Guo2016}.

In recent experiments on fluids, it was discovered that inhomogeneities in density $\rhoO$ and compressibility $\kapO$, introduced by a solute concentration field, can be acoustically relocated into stabilized configurations~\cite{Deshmukh2014, Augustsson2016}. In subsequent work~\cite{Karlsen2016, Karlsen2017}, we showed that fast-time-scale acoustics in such inhomogeneous fluids gives rise to a time-averaged acoustic force density acting on the fluid on the slower hydrodynamic time scale, and that this force density leads to the observed relocation and stabilization of the inhomogeneities. The experiments also indicated that boundary-driven streaming is suppressed in inhomogeneous fluids~\cite{Augustsson2016}, and we hypothesized that this hitherto unexplored phenomenon can be explained by the non-dissipative acoustic force density.

In this Letter, we investigate the above hypothesis by unifying the theories of acoustic streaming~\cite{Nyborg1953a, Nyborg1958, Lighthill1978, Riley2001} and the acoustic force density~\cite{Karlsen2016}. We verify analytically the limiting cases of the unified theory, and proceed to develop a full numerical model of boundary-driven acoustic streaming in inhomogeneous viscous fluids. The multiple-time-scale model describes the dynamics and interactions on both the fast acoustic time scale and the slow hydrodynamic time scale. We use the theory to simulate the evolution of the acoustic streaming, as an acoustically stabilized density profile evolves by diffusion and advection. We furthermore measure experimentally the evolution of the acoustic streaming in an inhomogeneous aqueous iodixanol solution in an ultrasound-activated glass-silicon microchannel. Our main findings are (i) that the competition between the boundary-induced streaming stresses and the inhomogeneity-induced acoustic force density introduces a dynamic length scale $\Delta$ of the streaming vortex size, (ii) that initially $\Delta \ll \Delta_\mathrm{hom} \sim \mathrm{min}\big\lbrace \lambda /8 , \, H/4 \big\rbrace$, where $\Delta_\mathrm{hom}$ is the characteristic vortex size in a homogeneous fluid set by the acoustic wavelength $\lambda$ or the channel height $H$, and (iii) that in the bulk farther than $\Delta$ from the boundaries, the streaming flow is suppressed. The vortex size $\Delta$ increases in time, as the inhomogeneity is smeared out by diffusion and advection, and the vortices eventually expand into the bulk, making the acoustic streaming pattern similar to that in a homogeneous fluid. These findings are rationalized by simple scaling arguments.

While our analysis of acoustic streaming in inhomogeneous fluids should be of considerable fundamental interest, the suppression of acoustic streaming furthermore has potential applications in nanoparticle manipulation and enrichment. Indeed, acoustic streaming has been a major show-stopper in the successful acoustophoretic manipulation of bioparticles such as exosomes, vira, and small bacteria~\cite{Antfolk2017}, the reason being the unfavorable scalings of the acoustic radiation force and the streaming-induced drag force with smaller particle sizes~\cite{Barnkob2012a, Muller2012}. Already, there have been attempts to suppress acoustic streaming using pulsed actuation~\cite{Hoyos2013, Muller2015}, or to engineer streaming patterns in special geometries that allow nano-particle manipulation~\cite{Antfolk2014, Mao2017, Collins2017}. In this work, we use the standard chip design sketched in~\figref{fig_01}, which allows the injection of a layered fluid creating a density gradient across the channel width~\cite{Augustsson2016, Karlsen2016}.

\begin{figure}[!t]
\centering
\includegraphics[width=0.95\columnwidth]{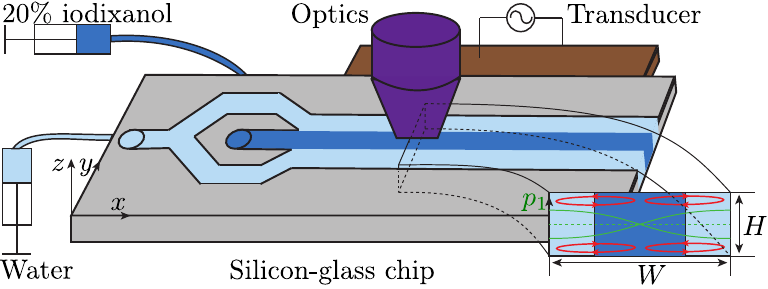}
\caption[]{\figlab{fig_01} (color online)
Sketch of the acoustofluidic silicon chip (grey) sealed with a glass lid, that allows the optical system (purple) to record the motion of the tracer beads (red trajectories) in the channel cross-section of width $W=375~\SIum$ and height $H=130~\SIum$. A 20\% iodixanol solution (dark blue) is injected in the center and laminated by pure water (light blue). The piezoelectric transducer (brown) excites the resonant half-wave pressure field $p_1$ (inset, green) at 2~MHz.}
\end{figure}

\textit{Separation of time scales.---} The foundation of the theory unifying the acoustic force density and the acoustic streaming is the separation of time scales between the fast acoustic time scale $t\sim 0.1~\upmu$s and the slow hydrodynamic time scale $\tau\sim 10~\SIms$~\cite{Karlsen2016}. Because of the large separation in time scales ($\tau \sim 10^5 t$), the acoustic fields can be solved for while keeping the hydrodynamic degrees of freedom fixed at each instance in time $\tau$. Assuming the system to be time-harmonically actuated at the angular frequency $\omega$, the density $\rho$ is thus written as
 \beq{PertExpansion}
 \rho = \rhoO(\rrr,\tau) + \rhoI(\rrr,\tau)\:\ee^{-\iot} .
 \eeq
Here, $\rhoO$ is the hydrodynamic density, and $\rhoI$ is the perturbation associated with the acoustic pressure and velocity fields $\pI$ and $\vvvI$.

\textit{Fast-time-scale acoustics.---} Using perturbation expansions of the form~\eqref{PertExpansion} for all fields in the equations for the conservation of fluid momentum and mass, one can show that the first-order equations for the acoustic perturbations $\vvvI$, $\pI$, and $\rhoI$, can be written as
 \bsub
 \eqlab{FirstOrderEqs}
 \bal
 - \ii \omega \rhoO \vvvI &= \divop\sigmabfI ,\\
 - \ii \omega \kapO \pI &= - \divop\vvvI , \\
 - \ii \omega \rhoO \kapO \pI &= - \ii \omega \rhoI + \vvvI \cdot \nablabf \rhoO .
 \eal
 \esub
Here, $\sigmabfI$ is the first-order of the fluid stress tensor, obtained by replacing $p$ by $\pI$ and $\vvv$ by $\vvvI$ in the usual expression for the fluid stress tensor $\sigmabf$~\cite{Karlsen2016}. The local speed of sound is $\cO = 1 / \sqrt{\rhoO\kapO}$.

In viscous acoustics, the oscillation velocity $\vvvI$ goes to zero at walls on the length scale set by $\delta= \sqrt{2\nu_0/\omega}$, with $\nuO=\etaO/\rhoO$, where $\nuO$ and $\etaO$ are the kinematic and dynamic viscosities, respectively. In water at 2 Mhz the boundary layer thickness is $\delta=0.4~\SIum$. It is within these narrow boundary layers, that the time-averaged stresses driving the streaming is generated. Neglecting viscosity in the acoustics,~\eqref{FirstOrderEqs} reduces to the standard wave equation in inhomogeneous media~\cite{Bergmann1946, Morse1986}.

\textit{Slow-time-scale dynamics.---} The fluid inhomogeneity is caused by a solute concentration field $s(\rrr,\tau)$, which is being transported on the slow timescale. This changes the hydrodynamic fluid density $\rhoO$, compressibility $\kapO$, and dynamic viscosity $\etaO$. Consequently,
 \beq{inhoms}
 \rhoO=\rhoO\big[s(\rrr,\tau)\big] , \ \kapO=\kapO\big[s(\rrr,\tau)\big] , \ \: \etaO=\etaO\big[s(\rrr,\tau)\big] .
 \eeq
The specific dependence of $\rhoO$, $\kapO$, and $\etaO$ on concentration $s$ of iodixanol are known from measurements~\cite{Augustsson2016, Karlsen2016}.

The hydrodynamics on the slow timescale $\tau$ is governed by the momentum-continuity and the mass-continuity equations for the fluid velocity $\vvv(\rrr,\tau)$ and pressure $p(\rrr,\tau)$, and the advection-diffusion equation for the concentration $s(\rrr,\tau)$ of the solute with diffusivity $D$,~\cite{Karlsen2016}
 \bsub
 \eqlab{DynamicsSlow}
 \bal
 \eqlab{NSSlow}
 \pp_\tau (\rhoO \vvv) &= \divop \big[ \sigmabf - \rhoO\vvv\vvv \big] + \fffac + \rhoO \gvec , \\
 \eqlab{ContSlow}
 \pp_\tau \rhoO &= - \divop \big( \rhoO \vvv \big) , \\
 \eqlab{DiffusionSlow}
 \pp_\tau s &= - \divop \big[ - D \nablabf s + \vvv s \big] .
 \eal
 \esub
Here, $\gvec$ is the acceleration due to gravity, $\sigmabf$ is the fluid stress tensor, and $\fffac$ is the acoustic force density.

All types of time-averaged acoustic flows, whether the classical Rayleigh and Eckart streaming flows~\cite{Nyborg1953a, Nyborg1958, Lighthill1978, Riley2001}, or the recently described relocation flows due to fluid inhomogeneities~\cite{Karlsen2016}, are driven by the divergence in the oscillation-time-averaged acoustic momentum-flux-density tensor $\avr{\Piac}$~\footnote{The time-averaging over one oscillation period $T=2\pi/\omega$ on the fast time scale $t$ is performed as $\avr{g}=\frac{1}{T}\int_0^T g(t) \mathrm{d}t$.}. A unifying formulation that spans all phenomena is achieved by defining $\fffac$ as,
 \beq{facDef}
 \fffac = -\divop\avr{\Piac} .
 \eeq
The oscillation-time-averaged acoustic momentum-flux-density tensor $\avr{\Piac}$ depends on products of the first-order acoustic fields. It is given by
 \bsub
 \eqlab{Piac}
 \bal
 \avr{\Piac} &= \avr{ \pii } \mathbf{1} + \avr{\rhoO\vvvI\vvvI} ,
 \\
 \avr{ \pii } &= \frac14 \kapO |\pI|^2 - \frac14 \rhoO |\vvvI|^2  .
 \eal
 \esub
In this expression, $\avr{\pii}$ is a local oscillation-time-averaged acoustic pressure. Importantly, in the general case of an inhomogeneous fluid, it depends on the solute concentration $s$, $\avr{\pii}=\avr{\pii(s)}$.

Combining \eqsref{facDef}{Piac}, we obtain the general expression for the acoustic force density valid for viscous inhomogeneous acoustics,
 \beq{facFull}
 \fffac = - \nablabf\avr{ \pii} - \divop\avr{\rhoO\vvvI\vvvI} .
 \eeq
This expression for $\fffac$ may be simplified in two special cases. First, in a viscous homogeneous fluid (with $s=0$), the local pressure $\avr{\pii}$ does not depend on $s$. As a result, the gradient term in $\fffac$ in~\eqref{facFull} can be absorbed into the pressure $p$ in the momentum equation~\eqrefnoEq{NSSlow} by redefining the pressure to be $\pti=p+\avr{\pii}$. Hence,
 \beq{facHom}
 \fffac^\mathrm{hom} = - \divop\avr{\rhoO\vvvI\vvvI} .
 \eeq
Indeed, this is how the driving terms are often presented in the classical~\cite{Nyborg1953a, Nyborg1958, Lighthill1978} and more recent~\cite{Muller2013, Riaud2017a, Lei2017} works on acoustic streaming, the governing equations being the time-independent versions of~\eqsref{NSSlow}{ContSlow}.

Second, considering inhomogeneous but inviscid acoustics, we recently demonstrated that~\eqref{facFull} yields~\cite{Karlsen2016},
 \beq{facInv}
 \fffac^\mathrm{invisc} = - \frac14 |\pI|^2 \nablabf\kapO - \frac14 |\vvvI|^2 \nablabf\rhoO .
 \eeq
It was further demonstrated that this non-dissipative force density is responsible for the slow-time-scale relocation of the fluid inhomogeneities into stable field-dependent configurations~\cite{Karlsen2016, Karlsen2017}.

In the context of boundary-driven acoustic streaming in an inhomogeneous fluid, the content of Eqs.~\eqrefnoEq{facFull}-\eqrefnoEq{facInv} is as follows: In the boundary layers, dissipation of acoustic energy leads to time-averaged stresses, confined on the length scale $\delta$, that causes boundary-driven streaming flows. However, in the presence of gradients in the density and compressibility of the fluid, a non-dissipative acoustic force density furthermore acts to stabilize certain inhomogeneity configurations, which may counteract the advective streaming flow. While~\eqsref{facHom}{facInv} demonstrate that these two force densities are present in viscous and inhomogeneous fluids, the two contributions cannot in general be separated analytically.

\textit{Numerical model in 2D.---} The dynamics in the 2D channel cross-section is solved numerically under a stop-flow condition with the initial condition sketched in \figref{fig_01} using a weak-form finite-element implementation in COMSOL Multiphysics~\cite{COMSOL52} with a regular grid of rectangular mesh elements~\footnote{The mesh-element size grows from a minimum of $\Delta h=0.1~\SIum$ in the boundary layers perpendicular to the walls to a maximum  of $\Delta h=1.3~\SIum$ in the bulk. The order of the Lagrange shape functions is quadratic for pressure and cubic for density and velocity.}. We use a segregated solver, which solves the time-dependent problem in two steps. In the first step, the fast-time-scale acoustics in the inhomogeneous media, as governed by~\eqref{FirstOrderEqs}, is solved while keeping the hydrodynamic degrees of freedom fixed on the timescale $\tau$. This allows evaluating the time-averaged acoustic force density $\fffac$ in~\eqref{facFull}. In the second step, the slow-time-scale dynamics governed by~\eqref{DynamicsSlow} is integrated in time using a generalized alpha solver with a damping parameter of 0.25, and a maximum time step of $\Delta \tau=7.5~\SIms$, while keeping the acoustic energy density fixed at $\Eac=50~\SIPa$~\footnote{To fix $\Eac$, which varies due to the evolution of $s(\rrr,\tau)$ and small shifts in resonance frequency, we adjust in each time step the sidewall actuation amplitude $d_0$~\cite{Muller2012}.}. This model extends our previous model work~\cite{Karlsen2016, Karlsen2017} by explicitly solving for the fast-time-scale viscous acoustics in the inhomogeneous medium, which is necessary for computation of the boundary-layer stresses that drive acoustic streaming.

\textit{Experimental method.---} The experiments were performed in a standard long straight microchannel of height $H=130~\SIum$ and width $W=375~\SIum$ in a silicon-glass chip with a piezoelectric transducer bonded underneath. A laminated flow of water and an aqueous 20\%-iodixanol solution (OptiPrep) was injected to form a concentration gradient with the denser fluid at the center, see~\figref{fig_01}. General Defocusing Particle Tracking (GDPT)~\cite{Barnkob2015} was used to record the motion of $1~\SIum$-diameter polystyrene tracer beads. The fluid streaming velocity was computed by subtracting the radiation-force-component from the bead veloicty~\cite{Barnkob2012a, Karlsen2015}. At time $\tau=0~\SIs$, the flow was stopped, and the GDPT measurements (10 fps) were conducted with the peak-to-peak voltage at the transducer input set to $2.5$~V, which corresponds to $\Eac=52~\SIPa$~\footnote{The estimate for $\Eac=(E_\mathrm{ac}^{20\%}+E_\mathrm{ac}^{0\%})/2$ is obtained from the measured energy densities $E_\mathrm{ac}^{20\%}$ and $E_\mathrm{ac}^{0\%}$ in homogeneous 20\% and 0\% iodixanol solutions.}, and a frequency sweep from 1.95 to 2.05~MHz in cycles of 10~ms, which yields a standing half-wave across the width~\cite{Manneberg2009a}. For each set of measurements, the particle motion was recorded for $160~\SIs$ to observe the evolution of the acoustic streaming. The experiment was repeated $N=16$ times to improve the statistics.

\begin{figure*}[!t]
\centering
\includegraphics[width=1.95\columnwidth]{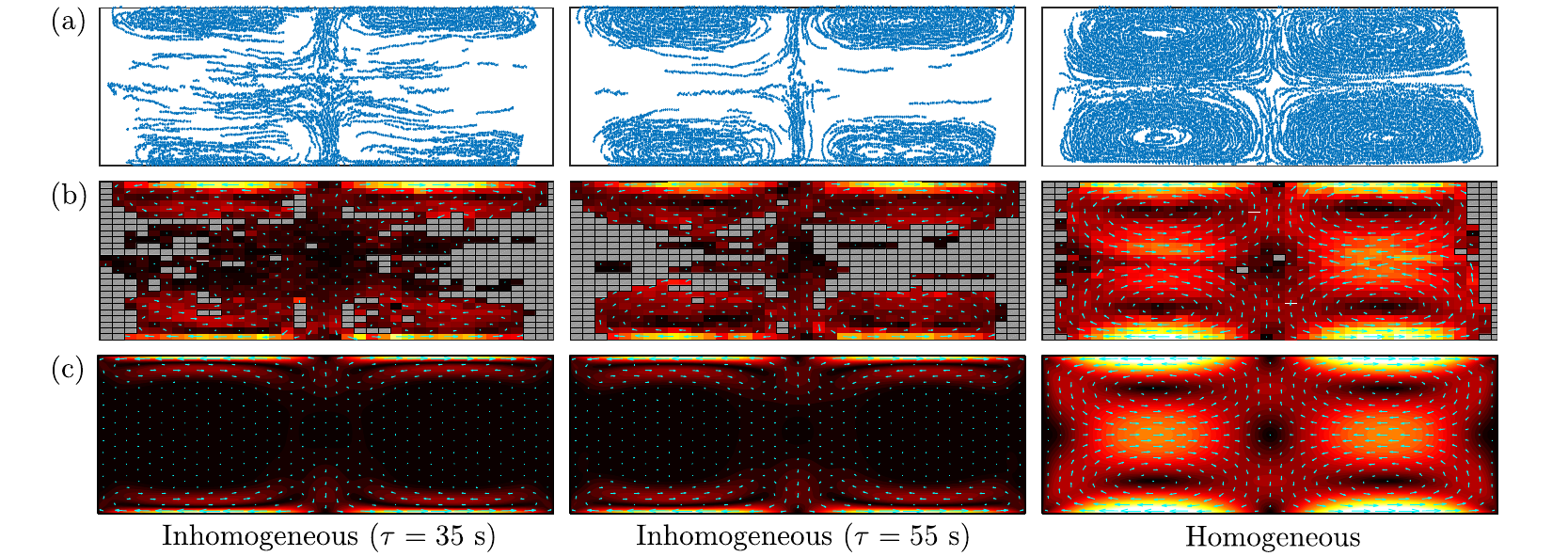}
\caption[]{\figlab{fig_02} (color online)
Acoustic streaming in the inhomogeneous fluid at $\tau=35~\SIs$ (1st column) and $\tau=55~\SIs$ (2nd column), measured experimentally in a symmetric $10~\SIs$ interval, and in the corresponding homogenized fluid (3rd column). (a) Experimental particle positions (blue points). (b) Experimental streaming velocity amplitude $|\vvv|$ ($0~\SIum/s$, black; $35~\SIum/s$, white) with the arrows (cyan) indicating the direction. Spatial bins with no data points are excluded (grey). (c) Simulated streaming velocity amplitude $|\vvv|$ ($0~\SIum/s$, black; $35~\SIum/s$, white) with the arrows (cyan) indicating the direction.}
\end{figure*}

\textit{Results.---} The experimental data and the simulation results for the acoustic streaming patterns in the channel cross-section are plotted for comparison in~\figref{fig_02}. The figure shows the inhomogeneous-fluid streaming at $\tau=35~\SIs$ (1st column), and $\tau=55~\SIs$ (2nd column), as well as the steady homogeneous-fluid streaming (3rd column). In the rows are (a) the raw experimental particle positions, (b) the grid-interpolated experimental velocity field, and (c) the simulated velocity field. The inhomogeneous-fluid streaming pattern evolves towards the homogeneous steady-state as diffusion (and, to a lesser extent, advection) diminishes the acoustically stabilized inhomogeneity, which has an initial 10\% excess mass density $\rhoHat_*$ at the center as compared to the sides~\footnote{$\rhoHat_*= \rhoO(0,\frac12 H) / \rhoO(\frac12 W,\frac12 H) - 1$, evaluated numerically.}. At $\tau=35~\SIs$ and $55~\SIs$, the excess mass density $\rhoHat_*$ has been reduced to 4\% and 2\%, respectively.

Evidently, the inhomogeneous-fluid streaming is initially confined close to the boundaries and suppressed in the bulk as compared to homogeneous-fluid streaming. To quantify this suppression of streaming, we define the vortex size $\Delta$ as the orthogonal distance from the boundary to the center of the streaming roll (where $\vvv=\zerovec$). In~\figref{fig_03}(a), the simulated vortex size $\Delta$ and the excess mass density $\rhoHat_*$ are plotted as functions of time. The vortex size $\Delta$ increases slowly in time, as the excess mass density $\rhoHat_*$ decreases by diffusion, until a transition occurs when a critically weak inhomogeneity is reached. At this point the streaming expands into the bulk and becomes similar to the streaming pattern in a homogeneous fluid. Figure~\ref{fig:fig_03}(a) shows that $\Delta$ and $\rhoHat_*$ are inversely related, supporting the hypothesis that the inhomogeneity-induced part of the acoustic force density [\eqref{facInv}] suppresses the boundary-driven streaming.

We may further assess the validity of the above-mentioned hypothesis by estimating $\Delta$ from a scaling argument. In the homogeneous-fluid case, the only relevant length scales are the channel dimensions $H$ and $W$ (the acoustic wavelength is $\lambda_0=2W$ by the assumption of a half-wave resonance). In the shallow-channel limit, the explicit analytical solution yields $\Delta_\mathrm{hom}= (1 - 1 / \sqrt{3}) (H/2)=28~\SIum$~\cite{Muller2013}. In a density-stratified medium, another length scale becomes relevant, namely the length scale $L_\rho$ of the gradient of the density $\nabla\rhoO \approx \rhoO/L_\rho$. Writing $\rhoO=\rhoOO[1+\rhoHat]$, the inhomogeneity-induced part of the acoustic force density [\eqref{facInv}] is of the order $f_\mathrm{ac} \approx \Eac \nabla \rhoHat$. We may then estimate $\Delta$ as the length scale on which the shear stress $\etaO \Lapl v_\mathrm{R} \approx \etaO v_\mathrm{R}/\Delta^2$, associated with the boundary-driven Rayleigh streaming velocity amplitude $v_\mathrm{R}=\frac32\Eac \rhoO^{-1} \cO^{-1}$~\cite{LordRayleigh1884}, is balanced by $f_\mathrm{ac}$. This scaling argument yields, using the early-time values $\rhoHat \approx 0.1$ and $L_\rho \approx W/2$,
 \beq{Lsup}
 \Delta
 \approx \sqrt{ \frac{3}{2} \frac{\nuO}{\cO} \frac{1}{|\nabla\rhoHat|}}
 \approx \sqrt{ \frac{3}{2} \frac{\nuO}{\cO} \frac{L_\rho}{\rhoHat}}
 \approx 2~\SIum .
 \eeq
This estimate for $\Delta$ is an order of magnitude smaller than $\Delta_\mathrm{hom}$, in good agreement with the experiments and simulations. It supports the hypothesis that $\Delta \ll \Delta_\mathrm{hom}$ due to the inhomogeneity-induced acoustic force density. Equation~\eqrefnoEq{Lsup} furthermore illustrates why the vortex size $\Delta$ grows in time; as time progresses, the inhomogeneity weakens by diffusion, i.e. $|\nabla\rhoHat|$ decreases, and consequently $\Delta$ grows.

\begin{figure}[!b]
\centering
\includegraphics[width=0.95\columnwidth]{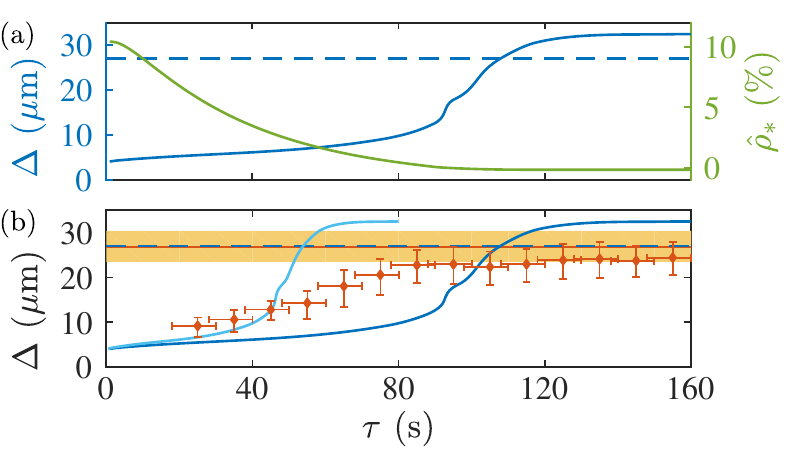}
\caption[]{\figlab{fig_03} (color online)
(a) Simulation results for the vortex size $\Delta$ (left axis, blue curve), $\Delta_\mathrm{hom}$ (left axis, dashed blue line), and the excess mass density $\rhoHat_*$ (right axis, green curve) as functions of time $\tau$. (b) Experimental results for $\Delta$ (red dots with errorbars) and $\Delta_\mathrm{hom}$ (red line with yellow error bar band), plotted with the simulation results for $\Delta$ (blue curve for unscaled time; light blue curve for rescaled time; see text) and $\Delta_\mathrm{hom}$ (dashed blue line) as functions of time $\tau$.}
\end{figure}

The time scale characterizing the growth of the vortex size $\Delta$ towards the value $\Delta_\mathrm{hom}$ is consequently set by diffusion. In the 2D simulation, where the diffusion is essentially 1D (across the width), the time scale of diffusion across one third of the channel width is $\tau_\mathrm{diff,1D}=(2D)^{-1}(W/3)^2=87~\SIs$. Figure~\ref{fig:fig_03}(a) shows a rapid transition in the simulated vortex size occurring around $\tau \approx 90~\SIs$, see also Supplementary Material \footnote{See Supplemental Material at [URL] for a simulation of 1-$\SImum$-diameter tracer particles for $0 < \tau <160$~s in the inhomogeneous acoustofluidic system.}. However, in the experiment we find that the transition occurs earlier and less rapid around $\tau \approx 60~\SIs$, see~\figref{fig_03}(b). Because axial variations in the acoustic field cannot be avoided in the experiment~\cite{Augustsson2011}, and because such variations lead to the loss of translational invariance, one can argue that axial concentration gradients render the diffusion 2D instead of 1D, which would halve the diffusion time, $\tau_\mathrm{diff,2D}=(4D)^{-1}(W/3)^2=43~\SIs$. Most likely, the effective diffusion in the experiment is in between the idealized 1D and 2D diffusion. In~\figref{fig_03}(b), the experimental data for the vortex size $\Delta$ is plotted as a function of time $\tau$, along with the simulation result for unscaled and rescaled time for 1D and 2D diffusion, respectively. The experimental data fall mostly between the two curves, and given that there are no free fitting parameters the agreement between theory and experiment is reasonable.

The 2D simulation successfully captures the essential physics of the experiment, including the initial suppression of streaming followed by the growth of the vortex size and the transition to a steady state. However, \figref{fig_03}(b) indicates that the simulation overestimates the long-time limit of $\Delta$. Interestingly, this is caused by an imperfect homogenization in 2D due to a delicate balance between the advective flows and the diffusive currents, leaving a slight over-concentration at the sidewalls (small negative $\rhoHat_*$; see \figref{fig_03}(a) at $\tau=160~\SIs$). Experimentally, however, the lack of perfect translational symmetry leads to homogeneous-fluid streaming at long time scales in agreement with  homogenized-fluid simulations, see \figref{fig_03}(b).

\textit{Conclusion.---} Theoretically, numerically, and experimentally, we have investigated the problem of acoustic streaming in inhomogeneous fluids with acoustically stabilized inhomogeneities. We have unified the theories of acoustic streaming and the acoustic force density, and developed a numerical model that simulates viscous inhomogeneous acoustics (the fast-time-scale dynamics) and the resulting flows due to the generalized acoustic force density (the slow-time-scale dynamics), allowing the interpretation of our experiments performed with aqueous iodixanol solutions in microchannels. We find that acoustic streaming is markedly different in homogeneous and inhomogeneous fluids as summarized by the main findings (i) - (iii) listed in the introduction. Our study is fundamental in scope, but the suppression of acoustic streaming in inhomogeneous fluids may enable ultrasound handling of nanoparticles in standard acoustophoretic chips.

%
%


%

\end{document}